\documentclass[lettersize,journal]{IEEEtran}
\usepackage{amsmath,amsfonts}
\usepackage{algorithmic}
\usepackage{array}
\usepackage{textcomp}
\usepackage{stfloats}
\usepackage{url}
\usepackage{verbatim}
\usepackage{graphicx}

\usepackage{cite}
\usepackage{amsmath,amssymb,amsfonts}
\usepackage{algorithmic}
\usepackage{graphicx}
\usepackage{textcomp}
\usepackage{amsmath}
\usepackage{amssymb}
\usepackage{latexsym}
\usepackage{CJK}
\usepackage{booktabs}

\usepackage{array}
\usepackage{multirow}
\usepackage{graphicx}
\usepackage{epstopdf}
\usepackage{caption}

\usepackage{makecell}
\usepackage{threeparttable}
\usepackage{bm}
\usepackage{subfig}
\usepackage{float}

\usepackage{algorithm}
\usepackage{algorithmic}
\usepackage{tabularx}
\usepackage{stfloats}
\usepackage{xcolor}
\usepackage{booktabs}

\def\BibTeX{{\rm B\kern-.05em{\sc i\kern-.025em b}\kern-.08em
    T\kern-.1667em\lower.7ex\hbox{E}\kern-.125emX}}
\usepackage{balance}
\begin{document}
\title{Nanoscale Reconfigurable Intelligent Surface Design and Performance Analysis for Terahertz Communications}
\author{ Xinying Ma, Zhi Chen, \IEEEmembership{Senior Member, IEEE}, and  Chongwen Huang, \IEEEmembership{Member, IEEE}

\thanks{This work was supported in part by the National Key Research and Development Program of China under Grant 2018YFB1801500, in part by the China Scholarship Council
(CSC) under Grant 202006070151.}

\thanks{X. Ma and Z. Chen are with National Key Laboratory of Science and Technology on Communications, University of Electronic Science and Technology of China, Chengdu 611731, China (e-mails: xymarger@126.com; chenzhi@uestc.edu.cn).

Chongwen Huang is with College of Information Science and Electronic Engineering, Zhejiang University, Hangzhou 310027, China (e-mails: chongwenhuang@zju.edu.cn).}
}

\markboth{Journal of \LaTeX\ Class Files,~Vol.~XX, No.~XX, June~2022}%
{How to Use the IEEEtran \LaTeX \ Templates}

\maketitle

\begin{abstract}
Terahertz (THz) communications have been envisioned as a promising enabler to provide ultra-high data transmission for sixth generation (6G) wireless networks. To tackle the blockage vulnerability brought by severe attenuation and poor diffraction of THz waves, a nanoscale reconfigurable intelligent surface (NRIS) is developed to smartly manipulate the propagation directions of incident THz waves. In this paper, the electric properties of the graphene are investigated by revealing the relationship between conductivity and applied voltages, and then an efficient hardware structure of electrically-controlled NRIS is designed based on Fabry-Perot resonance model. Particularly, the phase response of NRIS can be programmed up to 306.82 degrees. To analyze the hardware performance, we jointly design the passive and active beamforming for NRIS aided THz communication system. Particularly, an adaptive gradient descent (A-GD) algorithm is developed to optimize the phase shift matrix of NRIS by dynamically updating the step size during the iterative process. Finally, numerical results demonstrate the effectiveness of our designed hardware architecture as well as the developed algorithm.
\end{abstract}


\section{Introduction}
\IEEEPARstart{W}{ith} the continuous explosion growth of data traffic in wireless communications, sixth generation (6G) communication networks are expected to meet a great deal of pressing requirements in the near future \cite{introduction_01}. To meet the bandwidth-hungry applications (e.g., on-chip communication, virtual reality, kiosk downloading service, nanoscale localization and nanoscale biomedical communication),  terahertz (THz) frequency band (0.1-10 THz) has been regarded as a prospective alternative to provide large spectrum bandwidth and support ultra-high data transmission for 6G communication networks \cite{introduction_03}. However, there are still some imperative challenges existing in THz communications. On the one hand, due to the high path attenuation and strong molecular absorption effect experienced by THz waves, the transmission distance of THz communications is limited within a small area, and thus is not applicable for the practical communication scenarios \cite{introduction_08}, \cite{introduction_09}. On the other hand, THz waves at such a high frequency band undergo extremely poor diffraction and are easily blocked by the obstacles. To tackle this issue, the concept of nanoscale reconfigurable intelligent surface (NRIS) is newly proposed to mitigate blockage vulnerability and improve coverage capability \cite{introduction_10,introduction_12,introduction_13}. To be specific, the NRIS, which consists of a large number of passive reflecting elements is a kind of physical meta-surface and belongs to a special case of conventional RIS except for the extremely small physical size. Each reflecting element is capable of adjusting the phase shifts by using a smart central processor \cite{introduction_13-1}. In addition, NRIS is passive and lacks the active radio frequency (RF) chains, and thus is more energy-efficient compared with existing active devices, such as amplify-and-forward relaying \cite{introduction_17} and massive multiple-input multiple-output (MIMO). Therefore, the combination of NRIS and THz communications is worthy of further exploration.

Deploying NRISs in the THz communication system is essential, but some challenges also emerge accordingly. To realize reliable THz communications, the channel state information (CSI) acquisition is the primary mission before the data transmission begins. Different from conventional communication systems with active devices, the main difficulty of channel estimation in NRIS-enabled THz systems is that these reflecting elements are unable to execute the signal processing. Prominently, by leveraging the sparse features of THz MIMO channel, the work in \cite{introduction_21} converts the channel estimation problem into the sparse signal recovery problem, and a low complexity compressed sensing based channel estimation scheme is developed to realize the efficient signal reconstruction.
Once the CSI is acquired at the base station (BS) side or the mobile station (MS) side, the passive and active beamforming can be jointly designed for NRIS-enabled MIMO systems. For instance, the work in \cite{introduction_21-A} optimizes the transmit beamforming at the BS and the passive beamforming at the NRIS for the purpose of minimizing the total transmit power at the BS. To reap the benefits of multiple NRISs, the authors aim to solve a new cooperative multibeam multi-hop routing design problem and maximize the minimum received signal power among all users \cite{introduction_21-B, introduction_21-C}. Besides, the authors investigate the joint design of digital beamforming at the BS and analog beamforming at NRISs for the NRIS-enabled THz MIMO systems from the perspective of deep reinforcement learning \cite{introduction_21-D}.
In addition, other software research directions are also investigated extensively, such as energy efficiency optimization \cite{introduction_13, introduction_21-3}, data rate maximization \cite{introduction_22, introduction_23, introduction_24}, secure communication \cite{introduction_29}.
With regard to the hardware design of NRIS, many existing works pay attention to testing NRIS characteristics and implementing NRIS prototypes at THz frequency bands \cite{introduction_29-1,introduction_29-2,introduction_29-3}, but the algorithm design is not actually taken into consideration. Apart from these aforementioned research interests, the joint hardware design and hybrid beamforming for the NRIS-empowered THz MIMO communication system is still treated as an open problem.

In order to compensate for the research gap, the joint hardware design and hybrid beamforming optimization for the NRIS-enabled THz MIMO system is presented in this paper, \emph{ which is the first attempt to practically combine the hardware characteristics of the NRIS and the software design together}.
In particular, a novel graphene-based hardware structure of NRIS with a wide phase response range and a desired reflecting amplitude is designed. Considering the hardware features of NRIS, we develop a downlink NRIS-enabled THz MIMO system model and propose a gradient descent based method to optimize the phase shifts of NRIS.
Compared with the conventional MIMO system without NRIS \cite{introduction_32}, \cite{introduction_33}, the maximum rate optimization problem of the NRIS-enabled THz MIMO system involves multiple matrix variables, and thus is more sophisticated. The main contributions of this paper can be summarized as follows.

\begin{itemize}
  \item To begin with, we design a practical graphene-based NRIS hardware structure, where its  phase response can be controlled  up to 306.82 degrees, and the reflecting amplitude efficiency is more than 50{\%} at 1.6 THz. Furthermore, the design theory and working principle of the NRIS are  also provided. Then, the electric properties of the graphene are introduced by revealing the relationship between conductivity and applied voltage, which is the foundation of forming an electrically-controlled NRIS.

  \item Based on the hardware features of NRIS, we employ singular value decomposition (SVD) based method to obtain the optimally active beamforming at transceiver and propose an adaptive gradient descent (A-GD) algorithm to achieve the passive beamforming design for NRIS. Specifically, compared with conventional gradient descent (C-GD) algorithm with the fixed step size, the proposed A-GD algorithm turns out to be more efficient by dynamically updating the step size during the iterative process, and thus is able to realize a better achievable rate performance.

  \item Finally, simulation results demonstrate that our designed NRIS hardware structure with the practical phase response of 306.82 degrees achieves basically identical performance in contrast with the ideal phase response of 360 degrees. More importantly, with the assistance of NRIS, the achievable rate performance of our proposed A-GD algorithm greatly outstrips C-GD algorithm and the conventional THz systems without NRIS.
\end{itemize}

The rest of this paper is organized as follows. In Section II, the electric properties and hardware structure of graphene-based NRIS is presented. Section III describes the NRIS based software design for THz communications. Finally, the simulation results and conclusion are presented in Section IV and Section V, respectively.


\section{Graphene Based Hardware Design of NRIS}
In this section, an efficient graphene-based NRIS is designed that realizes a wide range of amplitude response and phase response. The key component of NIRS is the reflecting elements with sub-wavelength thickness,
and each reflecting element needs to be designed as a tunable resonant structure. In addition, by arranging these reflecting elements in a specific way, the whole NIRS is capable of achieving diverse functions, such as phase control, anomalous reflection and planar focusing. Finally, the detailed working principle and device performance can be presented in the following.

\subsection{Electric Properties of Graphene at THz Band}
In order to achieve the beam controllability, tunable components are embedded into the NRIS elements \cite{yaojia_01}. In general, semiconductor devices are extensively utilized at microwave or millimeter wave frequency band (e.g., varactor diode \cite{yaojia_02}, switching diode \cite{yaojia_03}). Since the physical size of each reflecting element at THz band is extremely small, the diodes and transistors can not be integrated into such a nanoscale structure. In this case, graphene is a kind of appropriate material to facilitate the NRIS with ultra-small size and tunable property.

Graphene is a two-dimensional material consisting of a single layer of carbon atoms. The conductivity of graphene can be altered through the applied voltage bias in a relatively wide range. Therefore, graphene provides various resonant states for each NRIS element. According to \cite{yaojia_08}, the conductivity of graphene at THz band can be written as
\begin{align}\label{yaojia_01}
\sigma {\rm{ = }}\frac{{2{e^2}}}{{\pi {\hbar ^2}}}{k_B}T \cdot \ln \left[ {2\cosh \left( {\frac{{{E_F}}}{{2{k_B}T}}} \right)} \right]\frac{i}{{\omega  + i{\tau ^{ - 1}}}},
\end{align}
where ${e}$ is the elementary charge; ${\hbar }$ is the reduced Planck constant; ${{{k_B}}}$ is the Boltzmann constant; ${T}$ is the temperature; ${{{E_F}}}$ is the Fermi level; ${\tau }$ is the relaxation time; and ${\omega }$ is the angular frequency, respectively. It can be concluded that at a certain frequency point, the conductivity is only determined by Fermi level \cite{yaojia_09}. Then we can get the following expression as
\begin{align}\label{yaojia_02}
\left| {{E_F}} \right| = \hbar {\nu _F}\sqrt {\pi {n_{d}}} ,
\end{align}
where ${{\nu _F}}$ is the Fermi velocity and ${n_{d}}$ is the carrier density which can be expressed as
\begin{align}\label{yaojia_03}
n_{d} = \sqrt {n_0^2 + \alpha_c {{\left| {{V_{{\rm{CNP}}}} - {V_g}} \right|}^2}} ,
\end{align}
where ${{n_0}}$ is the residual carrier density and ${\alpha_c }$ is capacitivity related to the electrode. Besides, ${V_{{\rm{CNP}}}}$ is the compensating voltage, and ${{V_g}}$ is the applied voltage \cite{yaojia_10}. In summary, the conductivity of graphene can be continuously changed by the applied voltages, which is the foundation of forming an electrically controlled NRIS.

\begin{figure}[!t]
\centering
\includegraphics[width=6.5cm]{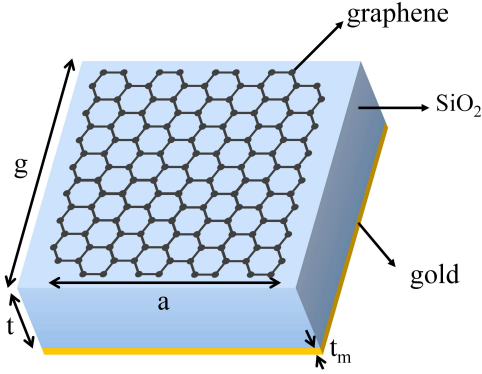}
\caption{Hardware structure of a nanoscale reflecting element that consists of graphene, quartz (substrate) and gold (ground plane) with ${a = 66{\mathop{\rm um}\nolimits} }$, ${g = 70{\mathop{\rm um}\nolimits} }$, ${t = 38{\mathop{\rm um}\nolimits} }$ and ${{t_m} = 1{\mathop{\rm um}\nolimits} }$.}\label{yaojia01}
\end{figure}

\subsection{Hardware Design of Graphene Based NRIS}
\begin{figure}[!t]
\centering
\subfloat[]{
\begin{minipage}[t]{1\linewidth}
\centering
\includegraphics[width=7cm]{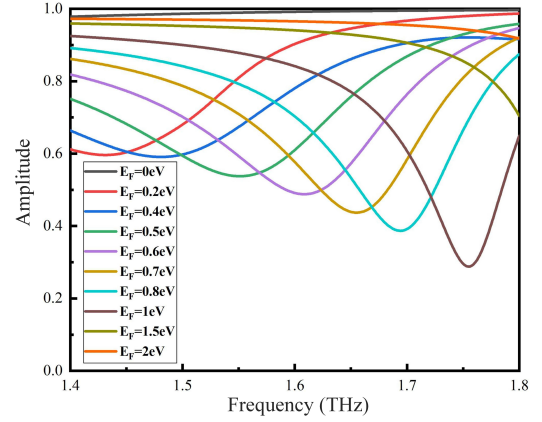}
\label{yaojia02a}
\end{minipage}%
}\\
\subfloat[]{
\begin{minipage}[t]{1\linewidth}
\centering
\includegraphics[width=7cm]{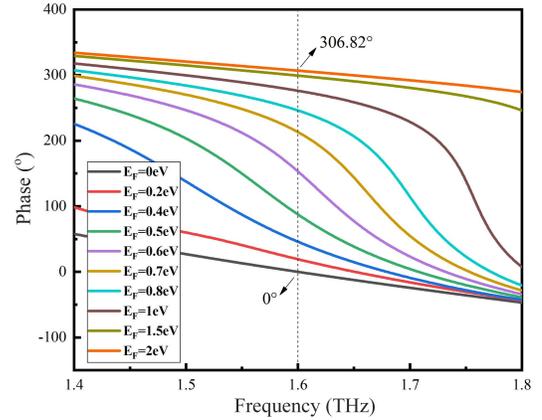}
\label{yaojia02b}
\end{minipage}
}\\
\subfloat[]{
\begin{minipage}[t]{1\linewidth}
\centering
\includegraphics[width=7.7cm]{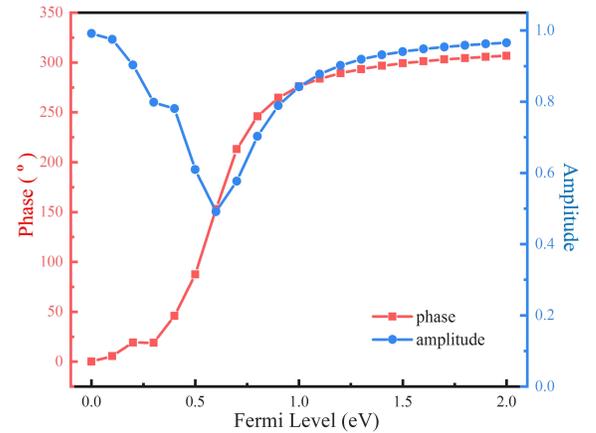}
\label{yaojia02c}
\end{minipage}%
}
\caption{Simulation results of the reflecting coefficient. (a) The response of reflecting amplitude; (b) The response of phase shift; (c) Phase response and reflecting amplitude versus Fermi level at 1.6 THz, where the phase response is ${\left[ {0^\circ ,306.82^\circ } \right]}$ with reflecting amplitude more than ${50\% }$.}\label{yaojia02}
\end{figure}

The EM responses of the reflecting elements play an important role in the hardware structure of the NRIS. Fig.\,\ref{yaojia01} shows a typical hardware design of a NRIS element, which can be divided into three parts from top to the bottom: the graphene layer, the substrate and the metallic ground plane. The resonance model of this NRIS architecture can be described as a Fabry-Perot cavity, where EM waves reflect back and forth between the top and the bottom surfaces. In addition, the resonance responses are caused by constructive or destructive interference of the multiple reflections \cite{yaojia_11}. In terms of such a reflecting element structure as shown in Fig.\,\ref{yaojia01}, the phase response based on \cite{yaojia_12} can be expressed as
\begin{align}\label{yaojia_04}
\varphi  = m\pi  - a{k_0}{\mathop{\rm Re}\nolimits} \left( {{n_{eff}}} \right),
\end{align}
where ${m}$ is an integer; ${a}$ is the width of graphene patch; ${{k_0}}$ is the wave number of free space; and ${{{n_{eff}}}}$ is the effective refraction index of the resonant structure, which is related to the effective permittivity ${{\varepsilon _{eff}}}$ of the graphene. In light of \cite{yaojia_13}, the parameter ${{\varepsilon _{eff}}}$ can be written as
\begin{align}\label{yaojia_05}
{\varepsilon _{eff}} = 1 + \frac{{i\sigma _c }}{{\omega {\varepsilon _0}{t_g}}},
\end{align}
where ${\sigma _c }$ is the conductivity and ${{{t_g}}}$ denotes the thickness of graphene. Combining (\ref{yaojia_04}) and (\ref{yaojia_05}), the phase response can be altered by the conductivity of graphene as well as the applied voltages.


The reflecting elements are simulated by leveraging the frequency domain solver in the simulation environment of CST Microwave Studio 2016. It is worth noting that a single reflecting element is unable to work since the miniature size causes the strong scattering. As a result, the boundary condition of the NRIS is set as `unit cell' to mimic the repeated arrangement of the NRIS elements. Fig.\,\ref{yaojia02} illustrates the reflecting coefficients from 1.4 THz to 1.8 THz with various Fermi levels. By combining Fig.\,\ref{yaojia02} (a) and Fig.\,\ref{yaojia02} (b), our designed NRIS performs relatively stable broadband characteristics. However, a narrowband working mode of the NRIS elements is selected in this paper where the center frequency is located at 1.6 THz.  Fig.\,\ref{yaojia02} (c) verifies that the amplitude efficiency of our designed reflecting element at 1.6 THz is more than ${50\% }$ and the phase response reaches to 306.82 degrees along with the chemical potential ranging from 0 ev to 2 eV. Moreover, the discrete phase shifts at 1.6 THz are also well-distributed with diverse Fermi levels, which lays the foundation for the bit quantization operation of phase shifts.

\begin{figure}[!t]
\centering
\includegraphics[width=8.8 cm]{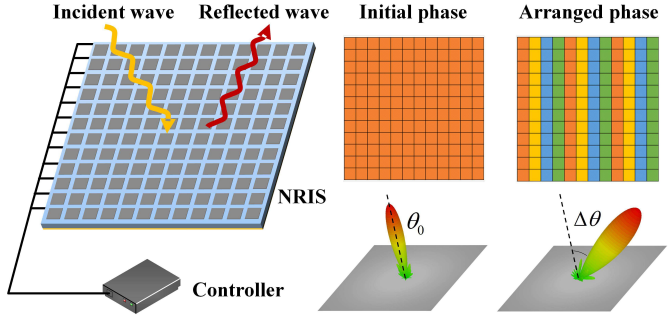}
\caption{Hardware design of the entire NRIS device that consists of a NRIS array and a controller. Various phase distributions represented by different color blocks are realized through applying different voltages to the nanoscale reflecting elements.}\label{yaojia03}
\end{figure}

Once the hardware structure of a single reflecting element is designed, the whole NRIS is able to be accomplished by arranging massive reflecting elements closely in an array structure, as shown in Fig.\,\ref{yaojia03}. Then, various beam steering functions can be realized by controlling the phase response of all the NRIS elements. But in practice, the appropriate number of the NRIS elements needs to be carefully selected, which can make a better tradeoff between performance and deployment cost. According to Fig.\,\ref{yaojia02} (c), any expected phase shift within the phase response range can be obtained via applying the voltages continuously. However, the continuous phase control for each NRIS element results in practical problems \cite{yaojia_14}, such as hardware complexity, power consumption, size limitation and the accuracy of phase control. To this end, the discrete phase shifts are considered as the hardware structure of NRIS.  We define the discrete phase set of each NRIS element as ${{\cal F} = \left\{ {0,{{{\varphi _{\max }}} \mathord{\left/ {\vphantom {{{\varphi _{\max }}} {{2^b}}}} \right. \kern-\nulldelimiterspace} {{2^b}}}, \cdots ,{{({2^b} - 1){\varphi _{\max }}} \mathord{\left/ {\vphantom {{({2^b} - 1){\varphi _{\max }}} {{2^b}}}} \right. \kern-\nulldelimiterspace} {{2^b}}}} \right\}}$, where ${\varphi _{\max }}$ is the maximum phase response and ${b}$ is the bit quantization number. Then, we define the reflecting amplitude set as ${{\cal A}}{\rm{ = \{ }}{\mu _1}, \cdots ,{\mu _{\left| {{\cal F}} \right|}}{\rm{\} }}$, where $\left| {{\cal F}} \right| = {2^b}$. From Fig.\,\ref{yaojia02} (c) we can note that  when the distribution of the discrete phase shifts is determined, $\cal A$ can be acquired accordingly. In other words, there is a fixed mapping relationship between phase shift and reflecting amplitude for a specific hardware structure of NRIS. Thus, the reflecting amplitude for each reflecting element can be further defined as $\bar \mu  = {{\left( {\sum\nolimits_{i = 1}^{|{{\cal F}}|} {{\mu _i}} } \right)} \mathord{\left/ {\vphantom {{\left( {\sum\nolimits_{i = 1}^{|{{\cal F}}|} {{\mu _i}} } \right)} {|{{\cal F}}|}}} \right. \kern-\nulldelimiterspace} {|{{\cal F}}|}}$, where the averaged amplitude $\bar \mu  \in \left[ {0.5,1} \right]$ is determined by parameter $b$ with the given NRIS structure. Specifically, considering these practical constraints, the maximum phase and amplitude response of NRIS are set as 306.82 degrees and 0.8, respectively.

\section{System Model and Beamforming Design}
In this section, we mainly introduce the NRIS-aided THz system model and present a joint passive and active beamforming design framework.

\subsection{System Model}
\begin{figure*}[!t]
\centering
\includegraphics[width=14 cm]{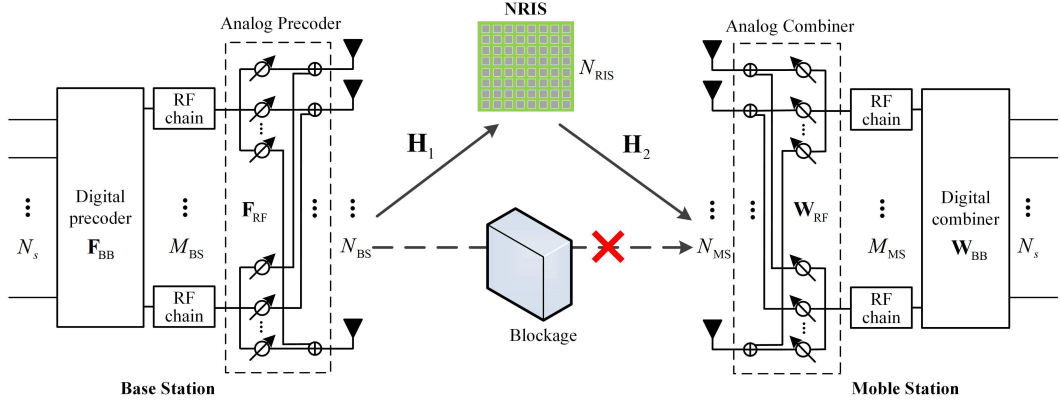}
\caption{Illustration of a NRIS-assisted downlink THz MIMO system.}\label{system01}
\end{figure*}


Considering a downlink THz MIMO system with hybrid beamforming architecture as shown in Fig. \,\ref{system01}, a BS employs ${{N_{{\mathop{\rm BS}\nolimits} }}}$ antennas to serve a MS equipped with ${{N_{{\mathop{\rm MS}\nolimits} }}}$ antennas. Since the LoS path between the BS and the MS is usually blocked by the obstacles, we suppose that the NRIS is installed to assist this THz communication link, which consists of ${{N_{{\mathop{\rm RIS}\nolimits} }}}$ passive reflecting elements. In addition, a controller that connects the BS and the NRIS is also required to realize the phase adjustment. Also, we assume that there are ${{M_{{\rm{BS}}}}}$ RF chains at BS side, and ${{M_{{\rm{MS}}}}}$ RF chains at MS side. Due to the serious power consumption of RF circuits, the number of antennas is larger than the number of the RF chains, i.e. ${{N_{{\rm{BS}}}} > {M_{{\rm{BS}}}}}$, ${{N_{{\rm{MS}}}} > {M_{{\rm{MS}}}}}$.
When the BS sends ${{N_s}}$ data streams ${{\bf{s}} \in {{\mathbb C}^{{N_s} \times 1}}}$ that satisfies ${\rm{E}}[{\bf{s}}{{\bf{s}}^H}] = \frac{1}{{{N_s}}}{{\bf{I}}_{{N_s}}}$, the MS employs ${{M_{{\rm{MS}}}}}$ RF chains to receive the processed signals. Thus, the received signal ${{{\bf{y}}_r} \in {{\mathbb C}^{{N_{{\mathop{s}\nolimits} }} \times 1}}}$ can be expressed as
\begin{align}\label{system_1}
{{\bf{y}}_r} = \sqrt \rho  {\bf{W}}_{{\mathop{\rm BB}\nolimits} }^H{\bf{W}}_{{\mathop{\rm RF}\nolimits} }^H{{\bf{H}}_2}{\bf{\Phi }}{{\bf{H}}_1}{{\bf{F}}_{{\mathop{\rm RF}\nolimits} }}{{\bf{F}}_{{\mathop{\rm BB}\nolimits} }}{\bf{s}} + {\bf{W}}_{{\mathop{\rm BB}\nolimits} }^H{\bf{W}}_{{\mathop{\rm RF}\nolimits} }^H{\bf{n}},
\end{align}
where ${\rho }$ is transmit power; ${{{\bf{H}}_1} \in {{\mathbb C}^{{N_{{\mathop{\rm RIS}\nolimits} }} \times {N_{{\mathop{\rm BS}\nolimits} }}}}}$ denotes BS-NRIS channel; ${{{\bf{H}}_2} \in {{\mathbb C}^{{N_{{\mathop{\rm MS}\nolimits} }} \times {N_{{\mathop{\rm RIS}\nolimits} }}}}}$ is NRIS-MS channel; ${{{\bf{F}}_{{\mathop{\rm RF}\nolimits} }} \in {{\mathbb C}^{{N_{{\mathop{\rm BS}\nolimits} }} \times {M_{{\mathop{\rm BS}\nolimits} }}}}}$ (${{{\bf{F}}_{{\mathop{\rm BB}\nolimits} }} \in {{\mathbb C}^{{M_{{\mathop{\rm BS}\nolimits} }} \times {N_s}}}}$) denotes the analog (digital) precoding matrix; ${{{\bf{W}}_{{\mathop{\rm RF}\nolimits} }} \in {{\mathbb C}^{{N_{{\mathop{\rm MS}\nolimits} }} \times {M_{{\mathop{\rm MS}\nolimits} }}}}}$ and ${{{\bf{W}}_{{\mathop{\rm BB}\nolimits} }} \in {{\mathbb C}^{{M_{{\mathop{\rm MS}\nolimits} }} \times {N_s}}}}$ are analog and digital combining matrices; ${{\bf{n}} \in {{\mathbb C}^{{N_{{\mathop{\rm MS}\nolimits} }} \times 1}}}$ represents the additive white Gaussian noise (AWGN) vector following the distribution of ${{\cal C}{\cal N}}\left( {{\bf{0}},{\delta ^2}{{\bf{I}}_{{N_{{\rm{MS}}}}}}} \right)$; and ${{\bf{\Phi }} = {\mathop{\rm diag}\nolimits} \left( {{{ { \mu {e^{j{\varphi _1}}}, \mu {e^{j{\varphi _2}}}, \cdots , \mu {e^{j{\varphi _{{N_{{\mathop{\rm RIS}\nolimits} }}}}}}} }}} \right)}$ is a diagonal matrix with the dimension of ${{N_{{\mathop{\rm RIS}\nolimits} }} \times {N_{{\mathop{\rm RIS}\nolimits} }}}$, respectively. Each entry $\left\{ { \mu {e^{j{\varphi _n}}}} \right\}_{n = 1}^{{N_{{\rm{RIS}}}}}$ of ${{\bf{\Phi }}}$ indicates the reflecting coefficient of a NRIS element, and is composed of the reflecting amplitude ${\mu}$ and the phase shift $\left\{ {{\varphi _n}} \right\}_{n = 1}^{{N_{{\rm{RIS}}}}}$. Both ${\mu}$ and ${\varphi _n}$ are closely related with the hardware structure of NRIS that is introduced in Section II.


\subsection{Channel Model}
The NRIS-enabled THz MIMO channel model contains ${{{\bf{H}}_1}}$, ${\bf{\Phi }}$ and ${{{\bf{H}}_2}}$, and the cascaded channel is denoted as ${{\bf{H}}_{\rm{e}}}{\rm{ = }}{{\bf{H}}_2}{\bf{\Phi }}{{\bf{H}}_1}$. We assume that both ${{{\bf{H}}_1}}$ and ${{{\bf{H}}_2}}$ consist of a LoS path and several non-line-of-sight (NLoS) paths, as we take the sparse nature of the THz channel into consideration. On the basic of geometric channel model \cite{system_01}, the BS-NRIS channel ${{{\bf{H}}_1}}$ can be written as
\begin{align}\label{system_3}
& {{\bf{H}}_1} = \sqrt {{N_{{\mathop{\rm BS}\nolimits} }}{N_{{\mathop{\rm RIS}\nolimits} }}} {\alpha _0}{{\bf{a}}_{{\mathop{\rm R}\nolimits} }}\left( {\theta _{{\mathop{\rm R}\nolimits} ,0}^1,\theta _{{\mathop{\rm R}\nolimits} ,0}^2} \right) {\bf{a}}_{{\mathop{\rm B}\nolimits} }^H\left( {\theta _{{\mathop{\rm B}\nolimits} ,0}^1,\theta _{{\mathop{\rm B}\nolimits} ,0}^2} \right) \notag \\
& + \sqrt {\frac{{{N_{{\mathop{\rm BS}\nolimits} }}{N_{{\mathop{\rm RIS}\nolimits} }}}}{L}} \sum\limits_{l = 1}^L {{\alpha _l}} {{\bf{a}}_{{\mathop{\rm R}\nolimits} }}\left( {\theta _{{\mathop{\rm R}\nolimits} ,l}^1,\theta _{{\mathop{\rm R}\nolimits} ,l}^2} \right){\bf{a}}_{{\mathop{\rm B}\nolimits} }^H\left( {\theta _{{\mathop{\rm B}\nolimits} ,l}^1,\theta _{{\mathop{\rm B}\nolimits} ,l}^2} \right),
\end{align}
where ${L}$ is the number of NLoS paths; ${\theta _{{\mathop{\rm R}\nolimits} }^1}$ $({\theta _{{\mathop{\rm R}\nolimits} }^2})$ denotes the angle of arrival (AoA) of ${{{\bf{H}}_1}}$ in the azimuth (elevation) domain; and ${\theta _{{\mathop{\rm B}\nolimits} }^1}$ $({\theta _{{\mathop{\rm B}\nolimits} }^2})$ denotes the angles of departure (AoD) of ${{{\bf{H}}_1}}$ in the azimuth (elevation) domain, respectively. Considering the large number of THz array antennas, the uniform planar array (UPA) structure is adopted as the array geometry. The normalized array response for the ${{N_x}{N_y}}$-element UPA on $xy$-plane can be expressed as
\begin{align}\label{system_4}
\begin{array}{l}
{{\bf{a}}_{{\rm{B}}}}{\rm{ = }}\frac{{\rm{1}}}{{\sqrt {{N_{{\rm{BS}}}}} }}\left[ {\cdots ,{e^{\frac{{j2\pi d}}{\lambda }\left( {p\sin \left( {\theta _{{\rm{B}}}^2} \right)\cos \left( {\theta _{{\rm{B}}}^1} \right) + q\cos \left( {\theta _{{\rm{B}}}^2} \right)} \right)}}, \cdots} \right]^T,
\end{array}
\end{align}
where $p \in \left[ {0,{N_x - 1}} \right]$, $q \in \left[ {0,{N_y - 1}} \right]$ and ${{N_x}{N_y} = {N_{{\mathop{\rm BS}\nolimits} }}}$, respectively. The spacing of THz array antenna is $d = {\lambda  \mathord{\left/ {\vphantom {\lambda  2}} \right. \kern-\nulldelimiterspace} 2}$ where $\lambda$ is the incident wavelength. Similar to (\ref{system_4}), ${{{\bf{a}}_{{\mathop{\rm R}\nolimits} }}\left( {\theta _{{\mathop{\rm R}\nolimits} }^1,\theta _{{\mathop{\rm R}\nolimits} }^2} \right)}$ also employs the UPA structure, but the spacing of the adjacent NRIS elements is the side length of each reflecting element. In addition, ${{\alpha _0}}$ is the LoS path gain of ${{{\bf{H}}_1}}$. As discussed in \cite{system_02}, ${{\alpha _0}}$ consists of spreading loss and molecular absorbing loss, which can be expressed as
\begin{equation}\label{system_5}
{\alpha _{\rm{0}}} = \frac{c}{{4\pi f{r_0}}} \cdot {e^{ - \frac{1}{2}\kappa \left( f \right){r_0}}} \cdot {e^{ - j2\pi f{\tau _{{\rm{Los}}}}}},
\end{equation}
where ${c}$ is the speed of light; ${r_0}$ is the straight distance between the BS and the MS; ${{\tau _{{\mathop{\rm Los}\nolimits} }} = {{r_0} \mathord{\left/ {\vphantom {r c}} \right.
 \kern-\nulldelimiterspace} c}}$ is the time-of-arrival of the LoS path; and ${{\kappa \left( f \right)}}$ is the molecular absorbing coefficient. In addition, the channel gain ${{\alpha _l}}$ for the ${l}$th reflected path based on \cite{system_02} and \cite{system_02-1} can be written as
\begin{align}\label{system_6}
{\alpha _l} = \frac{c (\xi \left( f \right))}{{4\pi  f  \left( {{r_1} + {r_2}} \right)}} \cdot {e^{ - \frac{1}{2}\kappa \left( f \right)\left( {{r_1} + {r_2}} \right) - j2\pi f{\tau _{{\mathop{\rm Ref}\nolimits} }}}},
\end{align}
where $\xi \left( f \right)$ is the reflection coefficient of the reflecting materials (e.g., concrete, plastic, glass); ${{{r_1}}}$ is the distance between the transmitter and the reflecting material; ${{{r_2}}}$ is the distance between the receiver and the reflecting material; and ${{\tau _{{\mathop{\rm Ref}\nolimits} }} = {\tau _{{\mathop{\rm Los}\nolimits} }} + {{\left( {{r_1} + {r_2} - r} \right)} \mathord{\left/ {\vphantom {{\left( {{r_1} + {r_2} - r} \right)} c}} \right. \kern-\nulldelimiterspace} c}}$ is the time-of-arrival of the reflected path, respectively. Besides, the channel characteristics of ${{{\bf{H}}_2}}$ are identical to ${{{\bf{H}}_1}}$, so we can generate ${{{\bf{H}}_2}}$ in the same way. We assume that the channel estimation problem has been extensively studied in \cite{introduction_21,system03}, and these effective channel estimation methods can be well leveraged in this paper.

\subsection{SVD-Based Active Beamforming Design}
In light of the system model and hardware architecture of NRIS, the achievable rate of the NRIS-aided THz MIMO system can be expressed as
\begin{align}\label{system_6}
&R = {\log _2}\left| {{{\bf{I}}_{{N_s}}} + \frac{\rho }{{{\delta ^2}{N_s}}}{{\left( {{{\bf{W}}^H}{\bf{W}}} \right)}^{ - 1}}{{\bf{W}}^H}} \right. \notag \\
&\quad\quad\quad\quad\quad\quad \left. { \times \left( {{{\bf{H}}_2}{\bf{\Phi }}{{\bf{H}}_1}} \right){\bf{F}}{{\bf{F}}^H}{{\left( {{{\bf{H}}_2}{\bf{\Phi }}{{\bf{H}}_1}} \right)}^H}{\bf{W}}} \right|,
\end{align}
where we define ${{\bf{F}} = {{\bf{F}}_{{\mathop{\rm RF}\nolimits} }}{{\bf{F}}_{{\mathop{\rm BB}\nolimits} }}}$ and ${{\bf{W}} = {{\bf{W}}_{{\mathop{\rm RF}\nolimits} }}{{\bf{W}}_{{\mathop{\rm BB}\nolimits} }}}$.
Then problem (\ref{system_6}) can be reformulated as
\begin{subequations}\label{system_7}
\begin{align}
& \left( {{{\bf{\Phi }}^{{\rm{opt }}}},{{\bf{W}}^{{\rm{opt }}}},{{\bf{F}}^{{\rm{opt }}}}} \right) = \mathop {\arg \max }\limits_{{\bf{\Phi }},{\bf{W}},{\bf{F}}} R \label{system70}\\
& \;\;\quad\quad\quad\quad{\rm{s.t.}}\;\; {\varphi _{\max }} = {306.82^{\rm{o}}}, \label{system7a} \\
& \;\;\quad\quad\quad\quad\quad\;\;\; {\bar \mu } = 0.8,\;\forall n = 1, \ldots ,{N_{{\rm{RIS}}}}, \label{system7b} \\
& \;\;\quad\quad\quad\quad\quad\;\;\; {\varphi _n} \in {{\cal F}},\;\forall n = 1, \ldots ,{N_{{\rm{RIS}}}}, \label{system7c} \\
& \;\;\quad\quad\quad\quad\quad\;\; \left\| {\bf{F}} \right\|_F^2 = {N_s},\label{system7d}
\end{align}
\end{subequations}
where (\ref{system7a}) is from our designed discrete phase response and (\ref{system7b}) stems from our designed reflecting amplitude. Due to the non-convex and discrete constraints, it is challenging to directly solve the problem (\ref{system_7}). Fortunately, one available way of settling such an optimization problem is to first design hybrid beamforming ${{\bf{F}}}$ and ${{\bf{W}}}$, and then optimize the phase shift matrix ${{\bf{\Phi }}}$, respectively.

Given a fixed ${{\bf{\Phi }}}$, the singular value decomposition (SVD) of the cascaded channel ${{\bf{H}}_{\rm{e}}}$ can be written as
\begin{align}\label{wenjie_3}
{{\bf{H}}_{\mathop{\rm e}\nolimits} } & = {\bf{U\Lambda }}{{\bf{V}}^H} \notag \\
& = \left[ {{{\bf{U}}_1},{{\bf{U}}_2}} \right]\left[ {\begin{array}{*{20}{c}}
{{{\bf{\Lambda }}_1}}&{\bf{0}}\\
{\bf{0}}&{{{\bf{\Lambda }}_2}}
\end{array}} \right]{\left[ {{{\bf{V}}_1},{{\bf{V}}_2}} \right]^H},
\end{align}
where ${{\bf{U}}}$ is a ${{N_{{\mathop{\rm MS}\nolimits} }} \times G}$ unitary matrix; ${{\bf{\Lambda }}}$ is a ${G \times G}$ dimensional matrix; ${{\bf{V}}}$ is a ${{N_{{\mathop{\rm BS}\nolimits} }} \times G}$ unitary matrix; ${{\bf{U}}_1}$ is a ${N_{{\rm{MS}}}} \times N_s$ submatrix of ${{\bf{U}} = \left[ {{{\bf{U}}_1},{{\bf{U}}_2}} \right]}$; and ${{\bf{V}}_1}$ is a ${N_{{\rm{BS}}}} \times N_s$ submatrix of ${{\bf{V}} = \left[ {{{\bf{V}}_1},{{\bf{V}}_2}} \right]}$, in which $G \buildrel \Delta \over = rank\left( {{{\bf{H}}_{\rm{e}}}} \right)$. Especially, ${{\bf{\Lambda }}_1}$ and ${{\bf{\Lambda }}_2}$ are $N_s \times N_s$ and $(G-N_s) \times (G-N_s)$ diagonal matrices with the singular values arranged in a decreasing order. According to \cite{introduction_33}, the optimal precoding matrix and combining matrix can be expressed as
\begin{align}\label{wenjie_3-1}
{{\bf{F}^{{\rm{opt}}}} = {{\bf{V}}_1} }, \;\;\; {{\bf{W}^{{\rm{opt}}}} = {{\bf{U}}_1}},
\end{align}
where the transmit power satisfies ${\left\| {\bf{F}^{{\rm{opt}}}} \right\|_F^2 = {N_s}}$.

\subsection{Proposed A-GD for Passive Beamforming Design}
Given the active beamforming matrices ${\bf{F}^{{\rm{opt}}}}$ and ${\bf{W}^{{\rm{opt}}}}$, the following target is to optimize the phase shift matrix ${\bf{\Phi }}$. In addition, the achievable rate ${R}$ in (\ref{system_6}) can be rewritten as
\begin{align}\label{wenjie_4}
R = {\log _2}\left| {{{\bf{I}}_{{N_s}}} + \frac{\rho }{{{\delta ^2}{N_s}}}{{\bf{\Lambda }}_1}{\bf{\Lambda }}_1^H} \right|.
\end{align}

Subsequently, (\ref{wenjie_4}) can be further simplified as
\begin{align}\label{wenjie_5}
R & = {\log _2}\left| {{{\bf{I}}_{{N_s}}} + \frac{\rho }{{{\delta ^2}{N_s}}}{\bf{\Lambda }}_1^2} \right| \notag \\
& \mathop  \le \limits^{(a)} {N_s}{\log _2}\left( {1 + \frac{\rho }{{{\delta ^2}{N_s}}}{\rm{tr}}\left( {{\bf{\Lambda }}_1^2} \right)} \right) \notag \\
& \mathop  \le \limits^{(b)} {N_s}{\log _2}\left( {1 + \frac{\rho }{{{\delta ^2}{N_s}}}{\rm{tr}}\left( {{{\bf{H}}_{\rm{e}}}{\bf{H}}_{\rm{e}}^H} \right)} \right),
\end{align}
where $(a)$ comes from Jensen's inequality and $(b)$ takes the mark of equality when $N_s = G$. Therefore, the optimization problem (\ref{system_7}) can be formulated as
\begin{subequations}\label{wenjie_5-1}
\begin{align}
{{\bf{\Phi }}^{{\rm{opt }}}} = & \mathop {\arg \max }\limits_{\bf{\Phi }} \;\; {\rm{tr}}\left( {{{\bf{H}}_{\rm{e}}}{\bf{H}}_{\rm{e}}^H} \right)\\
& \; {\rm{s.t.}}\;\; {\varphi _{\max }} = {306.82^{\rm{o}}},\\
& \;\;\;\;\;\;\; \left| {\bar \mu } \right| = 0.8,\;\forall n = 1, \ldots ,{N_{{\rm{RIS}}}},\\
& \;\;\;\;\;\;\;\; {\varphi _n} \in {{\cal F}},\;\forall n = 1, \ldots ,{N_{{\rm{RIS}}}}.
\end{align}
\end{subequations}

Nevertheless, it is worth noting that problem (\ref{wenjie_5-1}) is still a constrained optimization problem, since ${\bf{\Phi }}$ possesses discrete phase shifts and constant-magnitude entries. Let us define ${{\bm{\varphi }} \buildrel \Delta \over = \left[ {{{{{\varphi _1},{\varphi _2},...,{\varphi _{{N_{{\mathop{\rm RIS}\nolimits} }}}}} }}} \right]}$, and then consider ${{\bf{\Phi }}}$ as a function of ${\bm{\theta }}$ where ${\bm{\theta }} = {\bf{\Phi }}{{\bf{1}}_{{N_{{\rm{RIS}}}}}} = {[\bar \mu {e^{j{\varphi _1}}},\bar \mu {e^{j{\varphi _2}}}, \cdots ,\bar \mu {e^{j{\varphi _{{N_{{\rm{RIS}}}}}}}}]^T}$. Given this, we temporarily consider the continuous phase, and thus problem (\ref{wenjie_5-1}) can be reformulated as
\begin{subequations}\label{Ma11}
\begin{align}
& \mathop {\arg \min \;}\limits_{\bm{\theta }} f{\rm{(}}{\bm{\theta }}{\rm{) = }} - {{\bm{\theta }}^H}{\bf{D}}{\bm{\theta }}\\
& \; {\rm{s.t.}} \;\; {\varphi _{\max }} = {306.82^ \circ },\\
& \;\;\;\;\;\;\; \left| {{{\bm{\theta }}_n}} \right| = 0.8,\forall n = 1,2, \cdots ,{N_{{\rm{RIS}}}},
\end{align}
\end{subequations}
where $f{\rm{(}}{\bm{\theta }}{\rm{)}}$ is a compound function, ${\bf{D}} = {{{\bf{\hat D}}}^H}{\bf{\hat D}}$ and ${\bf{\hat D}} = [ \cdots ,{({\bf{H}}_1^T \otimes {{\bf{H}}_2})_{:,(n - 1){N_{{\rm{RIS}}}} + n}}, \cdots ]$, $\forall n = 1,2, \cdots ,{N_{{\rm{RIS}}}}$.


Note that problem (\ref{Ma11}) still suffers from constant-magnitude constraint due to the variable ${\bm{\theta }}$. To obtain an unconstrained problem, $f\left( {\bm{\theta}} \right)$ can be equivalently rewritten as
\begin{align}\label{Ma12}
f{\rm{(}}{\bm{\varphi }}{\rm{)}} & = -\sum\limits_{p = 1}^{{N_{{\rm{RIS}}}}} {\sum\limits_{q = 1}^{{N_{{\rm{RIS}}}}} {{\bm{\theta }}_p^H{{\bf{D}}_{p,q}}{{\bm{\theta }}_q}} } \notag \\
& = -{{\bar \mu }^2}\sum\limits_{p = 1}^{{N_{{\rm{RIS}}}}} {\sum\limits_{q = 1}^{{N_{{\rm{RIS}}}}} {{e^{ - j{\varphi _p}}}{{\bf{D}}_{p,q}}{e^{j{\varphi _q}}}}}.
\end{align}

Since ${\bf{D}} \in {{\mathbb C}^{{N_{{\rm{RIS}}}} \times {N_{{\rm{RIS}}}}}}$ is a positive-definite Hermitian matrix, we can get ${\bf{D}} = {{\bf{D}}^H}$. Given continuous phase shifts $\left\{ {{\varphi _n}} \right\}_{n = 1}^{{N_{{\rm{RIS}}}}}$, the ${n}$th element of the gradient vector ${{\nabla _{\bm{\varphi }}}{\mathop{f}\nolimits} \left( {\bm{\varphi }} \right)}$ can be calculated as
\begin{align}\label{Ma13}
\frac{{\partial f{\rm{(}}{\bm{\varphi }}{\rm{)}}}}{{\partial {\varphi _n}}} = & {{\bar \mu }^2}j{e^{ - j{\varphi _n}}}\sum\limits_{q = 1}^{{N_{{\rm{RIS}}}}} {{{\bf{D}}_{n,q}}{e^{j{\varphi _q}}}} \notag\\
& - {{\bar \mu }^2}j{e^{j{\varphi _n}}}\sum\limits_{p = 1}^{{N_{{\rm{RIS}}}}} {{{\bf{D}}_{p,n}}{e^{ - j{\varphi _p}}}}.
\end{align}

After calculating all the $\partial f({\bm{\varphi }})/\partial {\varphi _n},\forall n = 1,2, \cdots ,{N_{{\rm{RIS}}}}$, ${{\nabla _{\bm{\varphi }}}{\mathop{f}\nolimits} \left( {\bm{\varphi }} \right)}$ can be expressed as
\begin{align}\label{wenjie_19}
{\nabla _{\bm{\varphi }}}{\mathop{f}\nolimits} \left( {\bm{\varphi }} \right) = {\left[ {\frac{{\partial {\mathop{f}\nolimits} \left( {\bm{\varphi }} \right)}}{{\partial {\varphi _1}}},\frac{{\partial {\mathop{f}\nolimits} \left( {\bm{\varphi }} \right)}}{{\partial {\varphi _2}}}, \cdots ,\frac{{\partial {\mathop{f}\nolimits} \left( {\bm{\varphi }} \right)}}{{\partial {\varphi _{{N_{{\mathop{\rm RIS}\nolimits} }}}}}}} \right]^T}.
\end{align}

On basis of the gradient direction ${{\nabla _{\bm{\varphi }}}{\mathop{f}\nolimits} \left( {\bm{\varphi }} \right)}$, the objective function ${{\mathop{f}\nolimits} \left( {\bm{\varphi }} \right)}$ is able to descend by replacing ${{\bm{\varphi }}}$ with ${{\bm{\varphi }} - \lambda {\mathop{\rm diag}\nolimits} \left( {{\nabla _{\bm{\varphi }}}{\mathop{f}\nolimits} \left( {\bm{\varphi }} \right)} \right)}$, where ${\lambda }$ is the iterative step size. During the ${i}$th iteration, the updated ${{{\bm{\varphi }}^{\left( {i + 1} \right)}}}$ and the renewed ${{\mathop{f}\nolimits} \left( {{{\bm{\varphi }}^{i + 1}}} \right)}$ can be respectively written as
\begin{subequations}\label{Ma14}
\begin{align}
{{\bm{\varphi }}^{i + 1}} &= {{\bm{\varphi }}^i} - \lambda {\nabla _{\bm{\varphi }}}f\left( {{{\bm{\varphi }}^i}} \right) \label{Ma14A}\\
f\left( {{{\bm{\varphi }}^{i + 1}}} \right) &= f\left( {{{\bm{\varphi }}^i} - \lambda {\nabla _{\bm{\varphi }}}f( {{{\bm{\varphi }}^i}})} \right). \label{Ma14B}
\end{align}
\end{subequations}

Since the C-GD algorithm obtains the fixed step size ${{\lambda}}$ by simulation experiment \cite{introduction_10}, it suffers from high complexity and low efficiency. To this end, a novel A-GD algorithm is developed to determine the adaptive step size ${{\lambda ^i}}$. Based on above definition, (\ref{Ma14}) can be formulated as
\begin{align}\label{wenjie_22}
& f{({{\bm{\varphi }}^{i + 1}})} = - \sum\limits_{p = 1}^{{N_{{\rm{RIS}}}}} {\sum\limits_{q = 1}^{{N_{{\rm{RIS}}}}} {{{\left( {{\bm{\theta }}_p^{i + 1}} \right)}^H}{{\bf{D}}_{p,q}}{\bm{\theta }}_q^{i + 1}} } \notag\\
&  \quad\quad =  - {{\bar \mu }^2}\sum\limits_{p = 1}^{{N_{{\rm{RIS}}}}} {\sum\limits_{q = 1}^{{N_{{\rm{RIS}}}}} {{{\bf{D}}_{p,q}}{e^{j\left( {\varphi _q^i - \varphi _p^i} \right) + j\left( {\Delta \varphi _q^i - \Delta \varphi _p^i} \right)}}} },
\end{align}
where ${\Delta \varphi _n^i =  - {\lambda ^i}{{\partial {\mathop{f}\nolimits} \left( {{{\bm{\varphi }}^i}} \right)} \mathord{\left/ {\vphantom {{\partial {\mathop{f}\nolimits} \left( {{{\bf{\varphi }}^i}} \right)} {\partial \varphi _n^i}}} \right. \kern-\nulldelimiterspace} {\partial \varphi _n^i}}}$ for ${n = 1,2, \cdots ,{N_{{\mathop{\rm RIS}\nolimits} }}}$. Note that ${{\mathop{f}\nolimits} \left( {{{\bm{\varphi }}^{i + 1}}} \right)}$ is only determined by ${{\lambda ^i}}$ during $(i+1)$th iteration. Hence, the optimization problem with regard to ${{\lambda ^i}}$ can be expressed as
\begin{align}\label{Ma15}
{\lambda ^i} \approx \mathop {\arg \min }\limits_{{\lambda ^i}}  =  & - {{\bar \mu }^2}\sum\limits_{p = 1}^{{N_{{\rm{RIS}}}}} {\sum\limits_{q = 1}^{{N_{{\rm{RIS}}}}} {{{\bf{D}}_{p,q}}{e^{j\left( {\varphi _q^i - \varphi _p^i} \right)}}} } \notag\\
& \quad\quad\quad  \times {e^{j{\lambda ^i}\left( {\frac{{\partial f{\rm{(}}{{\bm{\varphi }}^i}{\rm{)}}}}{{\partial \varphi _p^i}} - \frac{{\partial f{\rm{(}}{{\bm{\varphi }}^i}{\rm{)}}}}{{\partial \varphi _q^i}}} \right)}}.
\end{align}

To smartly control ${{\lambda ^i}}$ during iterative process, we replace ${{e^{j{\lambda ^i}\left[ {{{\partial {\mathop{f}\nolimits} \left( {{{\bm{\varphi }}^i}} \right)} \mathord{\left/ {\vphantom {{\partial {\mathop{f}\nolimits} \left( {{{\bm{\varphi }}^i}} \right)} {\partial \varphi _p^i}}} \right. \kern-\nulldelimiterspace} {\partial \varphi _p^i}} - {{\partial {\mathop{f}\nolimits} \left( {{{\bm{\varphi }}^i}} \right)} \mathord{\left/ {\vphantom {{\partial {\mathop{f}\nolimits} \left( {{{\bf{\varphi }}^i}} \right)} {\partial \varphi _q^i}}} \right. \kern-\nulldelimiterspace} {\partial \varphi _q^i}}} \right]}}}$ by the second-order Taylor expansion formulation. Then, (\ref{Ma15}) can be approximated as
\begin{align}\label{Ma16}
{\lambda ^i} & \approx \mathop {\arg \min }\limits_{{\lambda ^i}} {\rm{ }} - {{\bar \mu }^2}\sum\limits_{p = 1}^{{N_{{\rm{RIS}}}}} {\sum\limits_{q = 1}^{{N_{{\rm{RIS}}}}} {{{\bf{D}}_{p,q}}{e^{j\left( {\varphi _q^i - \varphi _p^i} \right)}}} } \notag \\
& \quad\quad\quad\quad\quad \times \left( {1 + j{\lambda ^i}\Gamma _{p,q}^i + \frac{{{{(j{\lambda ^i}\Gamma _{p,q}^i)}^2}}}{2}} \right) \notag \\
& = \mathop {\arg \min }\limits_{{\lambda ^i}} {\rm{ }}{C_0} + {C_1}{\lambda ^i} + {C_2}{\lambda ^i},
\end{align}
where $\Gamma _{p,q}^i = {{{{\partial {\mathop{f}\nolimits} \left( {{{\bm{\varphi }}^i}} \right)} \mathord{\left/ {\vphantom {{\partial {\mathop{f}\nolimits} \left( {{{\bm{\varphi }}^i}} \right)} {\partial \varphi _p^i}}} \right. \kern-\nulldelimiterspace} {\partial \varphi _p^i}} - {{\partial {\mathop{f}\nolimits} \left( {{{\bm{\varphi }}^i}} \right)} \mathord{\left/ {\vphantom {{\partial {\mathop{f}\nolimits} \left( {{{\bf{\varphi }}^i}} \right)} {\partial \varphi _q^i}}} \right. \kern-\nulldelimiterspace} {\partial \varphi _q^i}}}}$.

\begin{algorithm}[!t]
	\caption{Proposed A-GD Algorithm}
	\begin{algorithmic}[1]
		\REQUIRE ${{{\bf{H}}_{\rm{1}}}}$, ${{{\bf{H}}_{\rm{2}}}}$, ${\rho }$, ${{N_s}}$, ${{\delta ^2}}$, ${{\cal F}}$, $b$, ${\varphi _{\max }}$, ${I}$,
		\STATE Initialize ${\bf{D}}$, ${{\bf{W}^{{\rm{opt}}}}}$,  ${{\bf{F}^{{\rm{opt}}}}}$, ${{\bf{\Phi }}} = \bar \mu {{\bf{I}}_{{N_{{\rm{RIS}}}} \times {N_{{\rm{RIS}}}}}}$,
                          \\ ${{{\bm{\varphi }}^0} = {{\bf{0}}_{{N_{{\mathop{\rm RIS}\nolimits} }} \times 1}}}$, ${{{\mathop{f}\nolimits} _{\max }} = 0}$, ${i = 0}$,
        \STATE \textbf{while} ${i \le I_{max}}$ \textbf{do}
        \STATE ${{}}$ ${{}}$ Calculate the gradient vector ${{\nabla _{\bm{\varphi }}}{\mathop{f}\nolimits} \left( {{{\bm{\varphi }}^i}} \right)}$ in (\ref{wenjie_19}),
        \STATE ${{}}$ ${{}}$ Calculate the step size ${{\lambda ^i}}$ according to (\ref{wenjie_29}),
        \STATE ${{}}$ ${{}}$ Update ${{{\bm{\varphi }}^{i + 1}} = {{\bm{\varphi }}^i} - {\lambda ^i} {{\nabla _{\bm{\varphi }}}{\mathop{f}\nolimits} \left( {{{\bm{\varphi }}^i}} \right)} }$,
        \STATE ${{}}$ ${{}}$ Update ${{\mathop{f}\nolimits} \left( {{{\bm{\varphi }}^{i + 1}}} \right) = {\mathop{f}\nolimits} \left( {{{\bm{\varphi }}^i} - {\lambda ^i} {{\nabla _{\bm{\varphi }}}{\mathop{f}\nolimits} \left( {{{\bm{\varphi }}^i}} \right)}} \right)}$,
        \STATE ${{}}$ ${{}}$ \textbf{if} ${{\mathop{f}\nolimits} \left( {{{\bm{\varphi }}^{i + 1}}} \right) > {{\mathop{f}\nolimits} _{\max }}}$ \textbf{do}
        \STATE ${{}}$ ${{}}$ ${{}}$ ${{}}$ ${{{\mathop{f}\nolimits} _{\max }} = {\mathop{f}\nolimits} \left( {{{\bm{\varphi }}^{i + 1}}} \right)}$, ${{{\bm{\varphi }}^{{\mathop{\rm opt}\nolimits} }} = {{\bm{\varphi }}^{i + 1}}}$,
        \STATE ${{}}$ ${{}}$ \textbf{end if}
        \STATE ${{}}$ ${{}}$ ${i = i + 1}$,
        \STATE \textbf{end while}
        \STATE Map each entry of ${{{\bm{\varphi }}^{{\mathop{\rm opt}\nolimits} }}}$ into discrete phase from ${{\cal F}}$,
        \ENSURE ${{{\bf{\Phi }}^{{\mathop{\rm opt}\nolimits} }}}$, ${R}$
	\end{algorithmic}
\end{algorithm}

It is worth noting that there are two different cases for the quadratic function in (\ref{Ma16}), including the positive value of term ${{C_2}}$ and the negative value of term ${{C_2}}$. For these two cases, ${{\lambda ^i}}$ can be calculated as
\begin{align}\label{wenjie_29}
{\lambda ^i} = \left\{ {\begin{array}{*{20}{c}}
{{{ - {C_1}} \mathord{\left/
 {\vphantom {{ - {C_1}} {\left( {2{C_2}} \right),{C_2} > 0}}} \right.
 \kern-\nulldelimiterspace} {\left( {2{C_2}} \right),{C_2} > 0}}}\\
{{{\left| {{C_1}} \right|} \mathord{\left/
 {\vphantom {{\left| {{C_1}} \right|} {\left| {{C_2}} \right|}}} \right.
 \kern-\nulldelimiterspace} {\left| {{C_2}} \right|}},\begin{array}{*{20}{c}}
{}
\end{array}{C_2} < 0}
\end{array}} \right.
\end{align}

The main steps of our proposed A-GD algorithm are illustrated in \textbf{Algorithm 1}.

\subsection{Complexity Analysis}
The main computational complexity of \textbf{Algorithm 1} comes from updating ${{\mathop{f}\nolimits} ( {{{\bm{\varphi }}^{i + 1}}})}$ in (\ref{Ma14}) and ${{\lambda ^i}}$ in (\ref{wenjie_29}). Specifically, the complexity of calculating (\ref{Ma14}) for each iteration can be expressed as ${\cal{O}} \left( 4N_{{\rm{RIS}}}^3 + 9N_{{\rm{RIS}}}^2 \right)$. For the sake of updating step size ${{\lambda ^i}}$, the complexity of executing (\ref{wenjie_29}) for each iteration is ${\cal{O}} \left( 2N_{{\rm{RIS}}}^3 + 8N_{{\rm{RIS}}}^2 \right)$. As a consequence, the total complexity of \textbf{Algorithm 1} can be given by
\begin{align}\label{com}
{\cal{O}} \left( {I_{max}\left( {6N_{{\rm{RIS}}}^3 + 17N_{{\rm{RIS}}}^2} \right)} \right).
\end{align}
where $I_{max}$ represents the maximum iterations of our proposed A-GD Algorithm.

\section{Simulation Results}

In this section, simulation results are provided to examine the effectiveness of joint hardware and algorithm design for NRIS-enabled THz system, including A-GD algorithm, C-GD algorithm, random phase algorithm and the conventional THz system without NRIS. In addition, the distances of BS-NRIS, NRIS-MS and BS-MS are set as ${{\bar r_0} = 10\;m}$, ${{\tilde r_0} = 20\;m}$, and ${{r_0} = 25\;m}$ respectively, which can commendably meet the indoor communication scenarios. The LoS path of BS-MS link is blocked by the obstacle, and thus the NLoS paths are assisted by NRIS. Considering the sparse nature of THz channel, we assume that ${{{\bf{H}}_1}}$ contains ${{L_1} = 3}$ propagation paths, e.g., one LoS path and two NLoS paths. More specifically, the complex gain of LoS path is generated based on (\ref{system_5}) and the complex gains of NLoS paths are computed by (\ref{system_6}). The molecular absorbing coefficient and the reflection coefficient of ceramic tile are set as ${\kappa (f) = 0.2}$ and ${\xi (f) = {10^{ - 6}}}$ \cite{system_02}. Similarly, the parameter settings of ${{{\bf{H}}_2}}$ are consistent with ${{{\bf{H}}_1}}$. We further set ${{N_{{\rm{BS}}}}{\rm{ = 512}}}$, ${{N_{{\rm{RIS}}}}{\rm{ = 256}}}$, ${{N_{{\rm{MS}}}}{\rm{ = 32}}}$, ${{M_{{\rm{BS}}}}}=6$ and ${{M_{{\rm{MS}}}}}=4$. Here we define the signal-to-noise ratio (SNR) as ${{\rm{SNR}} = \rho /{\sigma ^2}}$, and all simulation results are averaged over 1000 random channel realizations.

\subsection{Hardware Influence for THz System}

\begin{figure}[!t]
\centering
\includegraphics[width=7cm]{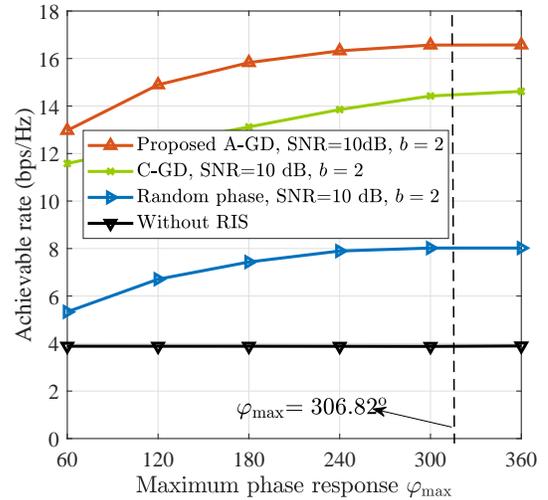}
\caption{Achievable rate comparisons  versus ${{\varphi _{\max }}}$.}\label{mage02}
\end{figure}

Fig. \,\ref{mage02} investigates the achievable rate with the increasing number of the maximum phase response ${{\varphi _{\max }}}$, which aims to validate the effectiveness of the proposed hardware architecture mentioned in Section II. From Fig. \,\ref{mage02} we can note that the achievable rates of the considered NRIS-enabled algorithms improve firstly and then converge to a fixed value. Remarkably, the achievable rate of our proposed A-GD  algorithm with ${{\varphi _{\max }} = {306.82^{\rm{o}}}}$ have already converged, and possess the same performance as the ideal case with ${{\varphi _{{\rm{ideal}}}} = {360^{\rm{o}}}}$. Compared with low phase response ${{\varphi _{\max }} = {60^{\rm{o}}}}$, the considered A-GD algorithm and the random phase scheme with ${{\varphi _{\max }} = {306.82^{\rm{o}}}}$ are able to realize the achievable rate enhancement of 3.59 bps/Hz and 2.68 bps/Hz, respectively. Hence, the numerical results reveal that our proposed graphene-based NRIS is an efficient hardware structure.

\begin{figure}[!t]
\centering
\includegraphics[width=7cm]{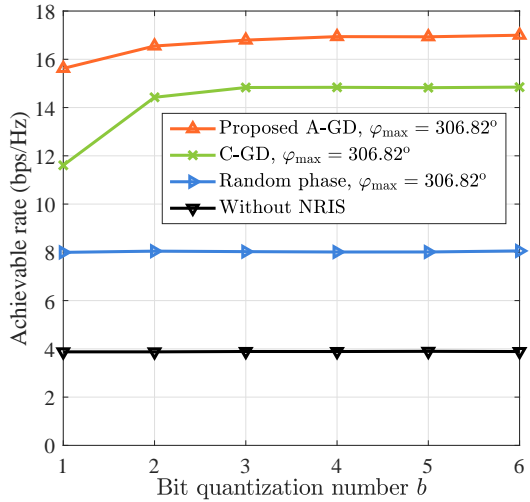}
\caption{Achievable rate comparisons versus ${b}$.}\label{mage04}
\end{figure}

Fig. \,\ref{mage04} provides the achievable rate performance versus different bit quantization values $b$. Fig. \,\ref{mage04} indicates that the achievable rates of the random phase scheme and the conventional THz system without NRIS are insensitive to $b$. Instead, the achievable rates of A-GD and AO algorithms are greatly affected by $b$. In the case of ${b = 1}$, our developed A-GD algorithm suffers from the obvious performance degradation due to the limited quantization precision of NRIS. In contrast with ${b = 2}$, our proposed A-GD algorithm with ${b = 1}$ endures about 0.93 bps/Hz performance penalty, respectively. Intriguingly, the achievable rates of our proposed algorithms with  ${b = 2}$ almost yields the similar performance compared with ${b \ge 3}$, indicating that ${b = 2}$ is sufficient to quantize the discrete phase for NRIS in practice.

\subsection{Performance Analysis for NRIS-Aided THz System}
\begin{figure}[!t]
\centering
\includegraphics[width=7cm]{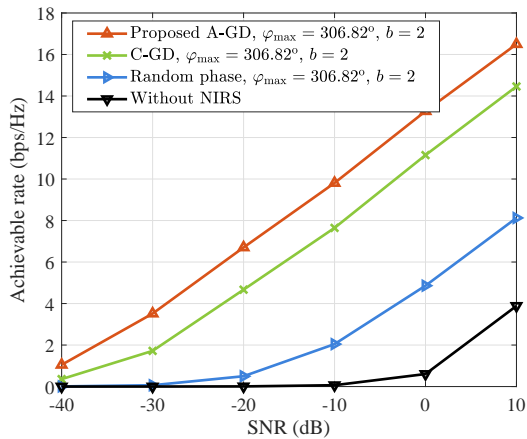}
\caption{Achievable rate comparisons versus SNR.}\label{mage01}
\end{figure}

\begin{figure}[!t]
\centering
\includegraphics[width=7cm]{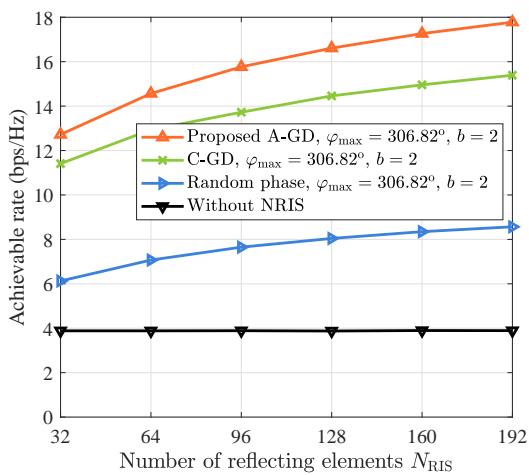}
\caption{Achievable rate comparisons versus ${{N_{{\rm{RIS}}}}}$.}\label{mage06}
\end{figure}

Fig. \,\ref{mage01} depicts the achievable rate versus diverse SNR in NRIS-aided THz MIMO system. As shown in Fig. \,\ref{mage01}, the achievable rate of the NRIS-aided THz system greatly outstrips the conventional THz system without NRIS that hardly meet the future communication requirements. Numerically, the performance gap between the random phase scheme and the THz system without NRIS is about 4.24 bps/Hz under the condition of SNR=10 dB. Meanwhile, Fig. \,\ref{mage01} also indicates that the achievable rate of our proposed A-GD algorithm is around 8.4 bps/Hz higher than the random phase scheme, respectively. More importantly, by dynamically adjusting the step size, our proposed A-GD algorithm is superior to the C-GD algorithm with fixed step size, which demonstrates that our proposed optimization algorithm can be employed to further enhance the achievable rate performance for the NRIS-aided THz MIMO systems.

Fig. \,\ref{mage06} discusses the achievable rate comparisons of the considered schemes versus the number of NRIS elements. From Fig. \,\ref{mage06}, we can note that the achievable rate of the THz system without NRIS case has the worst performance due to the lack of NRIS, and remains unchanged along with the diverse values of ${{N_{{\rm{RIS}}}}}$. In terms of these NRIS-aided algorithms, the achievable rate of our developed A-GD algorithm outstrips the C-GD algorithm and the random phase scheme. Moreover, when the number of reflecting elements increases, the performance gaps between A-GD algorithm and the random phase scheme become much larger. Specificall, our developed A-GD algorithm realizes around 9.21 bps/Hz performance improvement compared with the random phase scheme with ${{N_{{\rm{RIS}}}}{\rm{ = 192}}}$. Hence, NRIS can provides obvious performance gain for THz communications, and our proposed software design is capable of further improving the achievable rate. Last but not least, it should be pointed out that the achievable rates of NRIS-aided algorithms will converge to finite values even if the number of reflecting elements goes to infinity, which is caused by the power budget limitation existing in practical communications.

\section{Conclusion}
This paper jointly considered the joint hardware and software design for the NRIS-empowered THz MIMO communications. Considering the working principle of the NRIS, we primarily probed into the characteristics of the practical graphene-based NRIS. In light of these practical hardware constraints, the A-GD algorithm was developed to settle the passive beamforming design at NRIS and obtained much better performance than the C-GD algorithm. In the near future, our research work will concentrate on the practical measurements for the NRIS-empowered THz MIMO communication system.
%


\end{document}